\documentclass[a4paper,aps,twocolumn]{revtex4}
\RequirePackage[english]{babel}
\RequirePackage[latin1]{inputenc}
\RequirePackage[T1]{fontenc}
\RequirePackage{mathrsfs}
\RequirePackage{amsmath}
\RequirePackage{amssymb}
\RequirePackage{amsbsy}
\RequirePackage{bm}
\begin{document}
\title{\bf{A DISCUSSION ON THE MOST GENERAL TORSION-GRAVITY WITH ELECTRODYNAMICS FOR DIRAC SPINOR MATTER FIELDS}}
\author{Luca Fabbri}
\affiliation{INFN \& Dipartimento di Fisica, Universit\`{a} di Bologna, Via Irnerio 46, 40126 Bologna, ITALY}
\date{\today}
\begin{abstract}
We consider the most general torsional completion of gravitation together with electrodynamics for the Dirac spinorial material fields, and we show that consistency arguments constrain torsion to be completely antisymmetric and the dynamics to be parity-invariant and described by actions that are either least-order derivative or renormalizable.
\end{abstract}
\maketitle
\section*{INTRODUCTION}
In this paper, we will take into account the torsional completion of gravity, alongside to electrodynamics, as the fundamental quantities constituting the geometry that hosts Dirac spinor matter fields: the reason for doing this is that when the spacetime torsion is present beside the spacetime curvature, and beside the gauge curvature, the general underlying background is equipped with the possibility to couple the spin beside the energy, and beside the gauge currents, in what is the most extensive coupling that can be assigned to fermionic matter fields.

In general, torsion is decomposable in terms of three irreducible parts, and of these three parts the completely antisymmetric part is one enjoying a privileged position as it is discussed in \cite{Capozziello:2001mq,a-l,m-l,Fabbri:2006xq,Fabbri:2009se,Fabbri:2008rq,Fabbri:2009yc} and \cite{FV,Fabbri:2011kq}; and although in the presence of torsion parity-violating terms can be allowed in the action, parity-conservation tends to be restored at an effective level, as discussed in \cite{Fabbri:2013gza}: the property of the torsion to be completely antisymmetric and the feature of the dynamics to be parity-even evoke a peculiar character of the model which should be investigated in more depth.

From what may seem an unrelated perspective, when in theoretical physics we have to model a system we need a practical principle that could narrow all possibilities down to a few actual Lagrangians: so far as we can tell the only known way we have to do this is by invoking the assumption of renormalization: a Lagrangian is said to be renormalizable when at short scales a field's dynamics is not irrelevant compared to its interactions; however, if this were the case, it would mean that the field would tend to vanish, and not only this is not a contradiction, but also we already know that this does happen for some field, such as the gravitational field; moreover, although renormalization may keep a field safe at short distance where ultra-violet divergences are avoided, nevertheless that field would not be safe at large distances where infra-red divergences may occur, and there would be problems for massless fields, as it is well known for the case of the electrodynamic field. As it stands, such an assumption is quite arbitrary, and it would be wise either to justify it or to replace it in terms of a justified principle.

Although apparently disconnected, actually this problem is closely linked to the presence of torsion in gravity with electrodynamics; in fact, we will see that by assuming the presence of torsion beside curvature, and beside gauge fields, it is possible to show that consistency arguments imply torsion to be completely antisymmetric and the dynamics to be parity-even, and also they will imply the action to be renormalizable. There are problems that are usually solved by imposing some prescription of field quantization but which may also be addressed in terms of the non-linear potentials in the matter field equation as it is discussed in 
\cite{s-s,Fabbri:2010rw,Fabbri:2012ag,Fabbri:2013isa} and \cite{Fabbri:2014naa,Fabbri:2014zya}, and in this situation the possibility to have also renormalization related to the non-trivial geometry acquires additional interest.

In this paper we will investigate this possibility.
\section*{FOUNDATIONS}
The first thing we will specify is that we will work in the simplest space, that is the 
$(1\!+\!3)$-dimensional spacetime.

In such a spacetime, we will have both a differential structure and a metric structure, the former given in terms of the most general covariant derivative $D_{\mu}$ defined in terms of the most general connection, the latter given in terms of the most general metric $g_{\mu\nu}$ which is also used to raise and lower tensorial indices, and $D_{\alpha}g_{\mu\nu}\!=\!0$ specifies the compatibility of these two structures; the most general covariant derivative can be decomposed in terms of the simplest covariant derivative $\nabla_{\mu}$ defined in terms of the simplest connection, entirely written in terms of the metric $g_{\mu\nu}$ alone, and $\nabla_{\alpha}g_{\mu\nu}\!=\!0$ holds: such a decomposition is best seen in terms of their respective connections
\begin{eqnarray}
&\Gamma^{\alpha}_{\mu\nu}\!=\!\Lambda^{\alpha}_{\mu\nu}\!+\!K^{\alpha}_{\phantom{\alpha}\mu\nu} \label{connection}
\end{eqnarray}
in which we have that $\Gamma^{\alpha}_{\mu\nu}$ is the most general connection and $\Lambda^{\alpha}_{\mu\nu}\!=\!\frac{1}{2}g^{\rho\alpha}(\partial_{\mu}g_{\nu\rho}
\!+\!\partial_{\nu}g_{\mu\rho}\!-\!\partial_{\rho}g_{\mu\nu})$ is the simplest connection written in terms of the metric while we have that $K^{\alpha}_{\phantom{\alpha}\mu\nu}\!=\!
\frac{1}{2}(Q^{\alpha}_{\phantom{\alpha}\mu\nu}\!+\!Q_{\mu\nu}^{\phantom{\mu\nu}\alpha}
\!+\!Q_{\nu\mu}^{\phantom{\nu\mu}\alpha})$ is called contorsion and it is given in terms of the torsion, decomposed as
\begin{eqnarray}
&Q_{\rho\mu\nu}\!=\!\frac{1}{6}W^{\alpha}\varepsilon_{\alpha\rho\mu\nu}
\!+\!\frac{1}{3}\left(V_{\nu}g_{\rho\mu}\!-\!V_{\mu}g_{\rho\nu}\right)\!+\!T_{\rho\mu\nu}
\label{torsion}
\end{eqnarray}
where $T_{\rho\mu\nu}\!=\!Q_{\rho\mu\nu}\!-\!\frac{1}{6}W^{\alpha}\varepsilon_{\alpha\rho\mu\nu}
\!-\!\frac{1}{3}\left(V_{\nu}g_{\rho\mu}\!-\!V_{\mu}g_{\rho\nu}\right)$ is the non-completely antisymmetric irreducible tensorial part given in terms of $W^{\alpha}\!=\! Q_{\rho\mu\nu}\varepsilon^{\rho\mu\nu\alpha}$ as the axial vectorial part and $V_{\nu}\!=\!Q^{\rho}_{\phantom{\rho}\rho\nu}$ as the trace vectorial part of torsion.

In terms of the most general connection we define the Riemann curvature tensor $G_{\rho\eta\mu\nu}$ as usual, antisymmetric in the first and second couple of indices, with one independent contraction $G^{\rho}_{\phantom{\rho}\eta\rho\nu}\!=\!G_{\eta\nu}$ which itself has a contraction given by $G_{\eta\nu}g^{\eta\nu}\!=\!G$ and they are called Ricci curvature tensor and scalar; the simplest metric connection has the Riemann metric curvature tensor $R_{\sigma\eta\rho\nu}$ antisymmetric in both the first and second couple of indices and symmetric for a switch between the first and second couple of indices, with contraction $R^{\rho}_{\phantom{\rho}\eta\rho\nu}\!=\!R_{\eta\nu}$ itself with contraction $R_{\eta\nu}g^{\eta\nu}\!=\!R$ called Ricci metric curvature tensor and scalar: then we have the decomposition
\begin{eqnarray}
&\!\!\!\!G^{\rho}_{\phantom{\rho}\eta\mu\nu}\!\!=\!\!R^{\rho}_{\phantom{\rho}\eta\mu\nu}
\!+\!\!\nabla_{\mu}K^{\rho}_{\eta\nu}\!-\!\!\nabla_{\nu}K^{\rho}_{\eta\mu}
\!+\!\!K^{\rho}_{\sigma\mu}K^{\sigma}_{\eta\nu}\!\!-\!\!K^{\rho}_{\sigma\nu}K^{\sigma}_{\eta\mu}
\label{curvature}
\end{eqnarray}
in which the most general Riemann tensor is given in terms of the Riemann metric tensor and the contorsion.

Equivalently, it is possible to pass from this coordinate formalism into the Lorentz formalism, in which the covariant derivative $D_{\mu}$ is defined in terms of the most general spin-connection, and the metric is written according to the expression $g_{\alpha\nu}\!=\!\xi^{a}_{\alpha}\xi^{b}_{\nu}\eta_{ab}$ in terms of the orthonormal tetrad fields $\xi_{a}^{\sigma}$ and Minkowskian matrix $\eta_{ab}$ used to raise and lower Lorentz indices, and in Lorentz formalism we assume that $D_{\alpha}\xi_{\mu}^{j}\!=\!0$ and $D_{\alpha}\eta_{ij}\!=\!0$ hold: these two conditions of compatibility can be explicitly expressed as
\begin{eqnarray}
&\Gamma^{b}_{\phantom{b}j\mu}\!=\!\xi^{\alpha}_{j}\xi_{\rho}^{b} (\Gamma^{\rho}_{\phantom{\rho}\alpha\mu}+\xi_{\alpha}^{k}\partial_{\mu}\xi^{\rho}_{k})
\end{eqnarray}
and $\Gamma^{bj}_{\phantom{bj}\nu}\!=\!-\Gamma^{jb}_{\phantom{jb}\nu}$ showing that the spin-connection $\Gamma^{bj}_{\phantom{bj}\nu}$ is written in terms of the connection $\Gamma^{\alpha}_{\mu\nu}$ and the tetrad fields and that the spin-connection is antisymmetric in the two Lorentz indices compatibly with the requirement of Lorentz invariance, as we will see in the following.

In the equivalent Lorentz formalism, from the spin-connection we may define the Riemann curvature tensor in an analogous way: $G_{ab\mu\nu}\!=\!\xi^{\rho}_{a}\xi^{\eta}_{b}G_{\rho\eta\mu\nu}$ spells that the Riemann curvature in Lorentz formalism is the Riemann curvature in coordinate formalism after index renaming.

The passage from coordinate formalism to Lorentz formalism is important because in this way it is possible to convert the most general coordinate transformation law without any loss of generality into the special Lorentz transformation law, whose specific form makes it explicitly writable in terms of given representations, the real one but also the complex one, and when the representation is complex then fields have to be complex, and a new differential structure has to be given in terms of the gauge-covariant derivative $D_{\mu}$ defined in terms of the gauge-connection, as it is usually done in gauge theories.

From the gauge-connection alone it is possible to define the tensor given by $F_{\mu\nu}$ as the Maxwell strength, as usual. 

Of all Lorentz group's complex representations we will be interested in the simplest one, that is the one corresponding to the $\frac{1}{2}$-spin spinorial representation.

The differential structure is given by the most general spinorial covariant derivative $\boldsymbol{D}_{\mu}$ defined in terms of the most general spinorial connection and additionally we have to introduce the $\boldsymbol{\gamma}_{a}$ matrices belonging to the Clifford algebra $\{\boldsymbol{\gamma}_{i},\boldsymbol{\gamma}_{j}\}=2\boldsymbol{\mathbb{I}}\eta_{ij}$ from which we may define the matrices $\boldsymbol{\sigma}_{ij}\!=\!\frac{1}{4}[\boldsymbol{\gamma}_{i},\boldsymbol{\gamma}_{j}]$ as the antisymmetric matrices belonging to the complex Lorentz algebra, called spinorial algebra, and these matrices are such that the relationship $\{\boldsymbol{\gamma}_{a},\boldsymbol{\sigma}_{bc}\}\!=\!
i\varepsilon_{abcd} \boldsymbol{\pi}\boldsymbol{\gamma}^{d}$ implicitly defines the projection matrix $\boldsymbol{\pi}$ that will be used to define the left-handed and right-handed irreducible chiral decompositions of the spinor field, with the compatibility conditions now reading $\boldsymbol{D}_{\mu}\boldsymbol{\gamma}_{j}=0$ automatically: we have
\begin{eqnarray}
&\boldsymbol{\Gamma}_{\mu}=\frac{1}{2}\Gamma^{ab}_{\phantom{ab}\mu}\boldsymbol{\sigma}_{ab} +iqA_{\mu}\boldsymbol{\mathbb{I}}
\end{eqnarray}
showing that the most general spinorial connection can be written in terms of the Lorentz-valued spin-connection plus an abelian term that now may be identified with the gauge field described in terms of the gauge-connection.

And once again from the spinorial connection we define the spinorial version of the Riemann tensor in the usual manner: nevertheless, it is possible to see that 
\begin{eqnarray}
&\boldsymbol{G}_{\mu\nu}\!=\!\frac{1}{2}G^{ab}_{\phantom{ab}\mu\nu}\boldsymbol{\sigma}_{ab}
\!+\!iqF_{\mu\nu}\boldsymbol{\mathbb{I}}
\end{eqnarray}
with the spinorial curvature as the sum of the Lorentz-valued Riemann curvature and the Maxwell strength.

This introduction of the general setting that will constitute the underlying background of the paper served to settle the basic notation and conventions we will employ throughout the present article, although a more extensive exposition is given in \cite{Fabbri:2014zya} and references therein.
\subsection{Background Geometry}
Now that the background geometry has been defined, we may proceed to study the dynamics of the geometry and the matter it will contain, which is done by assigning the dynamical action or equivalently the Lagrangian.

In general, the action or the Lagrangian may contain up to an infinite number or terms, but this of course means that there will correspondingly be an infinite number of parameters to tune and in turn this diminishes the predictive power: to avoid this, one has to determine the Lagrangians by fixing them to a limited number of contributions, and this is done with some assumptions.

One such assumption is having the Lagrangian at the least-order derivative possible, that is having the contributions limited to those that have the lowest order of derivatives: from a theoretical perspective, lowest order of derivatives means fewest integration constants that will have to be chosen in looking for solutions; with this assumption, of all possible theories those that are picked are Einstein gravitation and Maxwell electrodynamics.

It would appear that this principle seems to possess a certain degree of viability, since it selects the two most successful theories ever established in physics; but on the other hand, there may be doubts cast on it for the fact that Einstein gravity does not have some of the features modern physics would demand, such as renormalizability.

As an alternative, one may then require the Lagrangian to be renormalizable, that is with contributions that are limited to those with $4$-dimension of mass in the kinetic term and $4$-dimension of mass and lower for interacting terms in general: theoretically, having up to $4$-dimension of mass means that when we scale the model as to reach higher energies, the kinetic terms will still be the most relevant contributions; with this assumption, of all possible theories those that are picked are various models of extended gravity and also Maxwell electrodynamics.

Despite the fact that the requirement of renormalizability demands for the replacement of Einstein gravity with one of its possible extensions, nevertheless we know of no such extension that is also free of problems, although for some higher-order theories of gravitation problems may be solved \cite{Bender2008/1, Bender2008/2}; nevertheless these results have never been proven for a general higher-order theory of gravity, and there is not a single problem-free extended model of gravity that is viable at present.

The assumption of least-order derivative gives rise to Einstein gravity, which is not renormalizable, and the requirement of renormalizability prompts the search for extended gravities, none of which is free of problems.

Here, we would like to introduce yet another requirement, one which might be more comprehensive than the two just discussed. We have already stressed that in general torsion couples to the spin in a similar way in which curvature couples to the energy, but there is no perfect symmetry between the roles of the energy and the spin, because while all fields have energy density not all fields have spin density: if the spin density tends to be smaller, torsion has to be smaller; if in the spin-torsion field equations there are curvature terms, they will not necessarily vanish, and the spin-torsion field equations will give rise to constraints on the curvature, not identically verified.

In this sense then, a theory with torsion may always be taken in the torsionless limit, but in this limit, it may be such that the curvature will be constrained in a way that is not always verified, and therefore we will speak about non-continuity: in reference \cite{Fabbri:2014kea} we have started to discuss the non-continuity of specific gravitational models. 

In the next section of the present paper, we will recall the concepts that have been first exposed in the above reference, and then fully deepen the investigation.
\subsubsection{Torsion-curvature crossed terms}
As it is clear from (\ref{connection}-\ref{torsion}), we may always separate metric and torsion and decompose the latter in three irreducible parts, and as it is clear from (\ref{curvature}), all of these parts will have mutual interactions between one another, and as a consequence, there is no loss of generality in treating all these quantities in their split form, and accounting for all interactions as well: we have then the Riemann metric curvature $R_{\alpha\mu\rho\sigma}$ and the three irreducible parts of torsion given by $T_{\rho\mu\nu}$, $W_{\alpha}$ and $V_{\nu}$ that have to be taken in all possible combinations, which have to be contracted in all indices configurations in order to give rise to all possible scalar terms; furthermore, it is known that for the curvature tensor we have the validity of the condition given by $R^{\rho}_{\phantom{\rho}\sigma\mu\nu} 
\!+\!R^{\rho}_{\phantom{\rho}\nu\sigma\mu} \!+\!R^{\rho}_{\phantom{\rho}\mu\nu\sigma}\!\equiv\!0$ and the Bianchi identities $\nabla_{\mu}R^{\nu}_{\phantom{\nu}\iota\sigma\rho} \!+\!
\nabla_{\sigma}R^{\nu}_{\phantom{\nu}\iota\rho\mu}
\!+\!\nabla_{\rho}R^{\nu}_{\phantom{\nu}\iota \mu \sigma}\!\equiv\!0$ while for the non-completely antisymmetric irreducible tensorial decomposition of the torsion tensor we have the constraint that is given by $T_{\rho\mu\nu}\!+\!T_{\mu\nu\rho}\!+\!T_{\nu\rho\mu}\!=\!0$ whose contraction is given according to $T^{\mu\nu\rho}\varepsilon_{\alpha\beta\mu\nu}\!=\!-\frac{1}{2}T^{\rho\mu\nu} \varepsilon_{\alpha\beta\mu\nu}$ itself with its own contraction $T^{\rho\mu\nu}\varepsilon_{\alpha\rho\mu\nu}\!=\!0$ by construction, and these are the set of identities needed to reduce all possible scalars to the core of independent scalar terms.

The resulting Lagrangian, obtained as the sum of these scalar terms, will be distinguish in three classes: the first class includes the least-order derivative models and it is the $2$-dimensional mass model; then there will be the renormalizable $4$-dimensional mass model; finally there will be all the remaining $n$-dimensional mass models.

Now, let us start the discussion about the possibility that a given Lagrangian yield field equations that display non-continuity: if in the Lagrangian there appears a term that is linear in the torsion tensor, then upon varying the Lagrangian the torsion-spin field equations will be formally written in the form of a combination of derivatives of torsion and possibly curvatures plus a spurious term without any torsion: in the limit where both spin density and torsion tend to vanish, the spin-torsion field equations remain in the form of the spurious term equals to zero, which is a constraint that is in general not necessarily verified; consequently, if we want no such circumstance, we must have a spin-torsion field equation that contains no spurious term; hence in the Lagrangian there must be no linear torsion term whatsoever.

To begin our analysis, let us consider the least-order derivative model: the $2$-dimensional mass model is characterized by a Lagrangian that can only contain one curvature and five square-torsion terms according to
\begin{eqnarray}
\nonumber
&\mathcal{L}\!=\!R\!+\!AT_{\rho\mu\nu}T^{\rho\mu\nu}
\!+\!BW_{\nu}W^{\nu}\!+\!CV_{\mu}V^{\mu}+\\
&+LT^{\rho}_{\phantom{\rho}\mu\nu}T_{\rho\eta\alpha}\varepsilon^{\mu\nu\eta\alpha}
\!+\!MW_{\mu}V^{\mu}
\end{eqnarray}
where the Newton constant is normalized to unity and with five torsional constants; terms that are linear in torsion are only divergences of vectorial parts of torsion 
\begin{eqnarray}
&\Delta \mathcal{L}\!=\!U\nabla_{\mu}W^{\mu}\!+\!Z\nabla_{\mu}V^{\mu}
\end{eqnarray}
and therefore dropped as irrelevant. Consequently we have that in the most general case continuity is ensured.

The following step consists in proceeding to the analysis of the renormalizable model: the $4$-dimensional mass model is characterized by a Lagrangian that can only contain squared-curvature, quartic-torsion, products of curvatures and squared-torsion, derivatives of cubic-torsion, second-derivatives of squared-torsion, and derivatives of products between curvature and torsion; terms that are linear in torsion are the derivatives of products between curvature and torsion, and it is possible to see that the independent contractions are given according to
\begin{eqnarray}
\nonumber
&\Delta \mathcal{L}
\!=\!J\nabla_{\alpha}R_{\beta\rho}T^{\rho}_{\phantom{\rho}\mu\nu}\varepsilon^{\alpha\beta\mu\nu}
\!+\!K\nabla_{\alpha}RW^{\alpha}+\\
&+E\nabla_{\nu}R_{\mu\rho}T^{\rho\mu\nu}
\!+\!F\nabla_{\alpha}RV^{\alpha}
\end{eqnarray}
and nothing else at all. In order to maintain continuity with no constraint on the curvatures, this contribution must disappear, and because of the independence of its terms, each single term must vanish; we would also like to avoid arbitrary tunings of the parameters, so that we are not going to require their vanishing unless some principle justifies this assumption: parity-evenness may be invoked to have $J$ and $K$ equal to zero but there is no principle for which $E$ or $F$ should be equal to zero and thus we have to ask that $T_{\alpha\mu\nu}$ and $V_{\alpha}$ vanish. When this requirement is assumed, we acknowledge that continuity is preserved in the case in which no tuning is assumed when we require parity-evenness and the presence of only the axial vectorial part of torsion: an axial vectorial part of torsion and parity-evenness are together necessary and sufficient conditions for the most general model to be continuous.

The case of further models is easy: the $n$-dimensional mass models for $n$ that is larger than $4$ are characterized by Lagrangians with all terms; but then terms that are given as the derivative of a curvature times another curvature times the axial vectorial part of torsion such as for example
$\nabla_{\rho}R_{\alpha\nu}R_{\pi}^{\phantom{\pi}\nu}W_{\kappa}\varepsilon^{\kappa\rho\alpha\pi}$ are parity-even and yet they do no vanish unless we require the supplementary constraint $W_{\iota}\!=\!0$ to hold. What this means is that for continuity to be preserved if no tuning is assumed then beside parity-evenness the presence of the axial vectorial part of torsion alone is no longer enough and one has to require that the axial vectorial part of torsion vanishes too, that is one has to require the vanishing of torsion, which is against the possibility to provide the coupling to spin also for systems of fermions; hence there is no way in which the most general model may be continuous.

We may summarize our results: $2$-dimensional mass models are continuous; $4$-dimensional mass models are continuous if and only if parity-invariant and with axial vector torsion; $n$-dimensional mass models are not continuous. We have two models: the least-order derivative Lagrangian and the one with axial vector torsion and parity-invariance for renormalizable Lagrangians.
\subsubsection{Abelian gauge fields}
A first point that needs to be clarified is the fact that in presence of torsion there is a generalization of the covariant derivatives of tensors that might in principle create problems in electrodynamics: the definition of the Maxwell tensor is given as the strength of the gauge-connection $F_{\mu\nu}\!=\!\partial_{\mu}A_{\nu}\!-\!\partial_{\nu}A_{\mu}$ and, if we see this definition as the curl of the gauge-connection then we should take the curl of the most general covariant derivatives given by $F_{\mu\nu}\!=\!D_{\mu}A_{\nu}\!-\!D_{\nu}A_{\mu}\!=\!
\partial_{\mu}A_{\nu}\!-\!\partial_{\nu}A_{\mu}\!+\!Q^{\rho}_{\phantom{\rho}\mu\nu}A_{\rho}$ which is not gauge invariant precisely because of the presence of the torsion tensor; but although the strength happens to coincide with the curl of the gauge-connection nevertheless the correct interpretation is to see the strength as the commutator of the gauge-covariant derivatives, in terms of which the most general definition of strength is exactly the one given 
by $F_{\mu\nu}\!=\!\partial_{\mu}A_{\nu}\!-\!\partial_{\nu}A_{\mu}$ because none of its generalizations would maintain $F_{\mu\nu}$ as to be the commutator of the gauge-covariant derivatives. Hence, Cauchy identities $\nabla_{\mu}F_{\sigma\rho}\!+\!\nabla_{\sigma}F_{\rho\mu}
\!+\!\nabla_{\rho}F_{\mu\sigma}\!\equiv\!0$ are unchanged, and similarly as above they will be needed to reduce all possible scalars to the core of independent scalar terms.

The absence of kinematic mixing between torsion and gauge curvature does not mean that there is no dynamical interaction between torsion-gravity and electrodynamics, and to show that there can be torsional interactions in electrodynamics we study the most general model for electrodynamics: the least-order derivative model is renormalizable, and it is the $4$-dimensional mass model.

The construction of the Lagrangian will have to adhere to the requisite of continuity; therefore, we require the Lagrangian to contain no linear torsion term.

As it was anticipated, the least-order derivative is the renormalizable model: the $4$-dimensional mass model is characterized by a Lagrangian that can only contain two strengths, products of one strength and two irreducible parts of torsion, and derivatives of strength times one irreducible part of torsion; terms that are linear in torsion are the derivatives of strength times one irreducible part of torsion, whose independent contractions are given by
\begin{eqnarray}
\nonumber
&\Delta\mathcal{L}
\!=\!J\nabla_{\alpha}F^{\mu\nu}T^{\alpha\rho\sigma}\varepsilon_{\mu\nu\rho\sigma}
\!+\!K\nabla_{\alpha}F^{\alpha\nu}W_{\nu}+\\
&+E\nabla_{\alpha}F_{\mu\nu}T^{\alpha\mu\nu}
\!+\!F\nabla_{\alpha}F^{\alpha\nu}V_{\nu}
\end{eqnarray}
and nothing else. As before, this contribution disappears when each single term vanishes: again, parity-evenness may be invoked to set $J$ and $K$ equal to zero but we have to ask that $T_{\alpha\mu\nu}$ and $V_{\alpha}$ vanish. Thus, continuity is preserved in the case in which no tuning is assumed if we require parity-evenness and the presence of only the axial vectorial part of torsion: an axial vectorial part of torsion and parity-evenness are together necessary and sufficient conditions for the most general model to be continuous.

So torsion does not affect electrodynamics, although electrodynamics has left a mark on torsion as the electrodynamic $4$-dimensional mass model is continuous when electrodynamics is invariant under parity and torsion is completely antisymmetric. There is one model, having completely antisymmetric torsion and parity-evenness for least-order derivative renormalizable Lagrangians.
\subsection{Material Content}
So far we have extensively discussed and thoroughly investigated the theory of torsional-gravitation with electrodynamics, and next and last step is that of introducing the general Lagrangian for Dirac spinorial matter fields.
\subsubsection{Spinor fields}
As we may always separate the left-handed and right-handed semi-spinorial chiral projections, it is easier to start with them: the left-handed and right-handed semi-spinors are $\psi_{L}\!=\!
\frac{1}{2}(\mathbb{I}\!-\!\boldsymbol{\pi})\psi$ and $\psi_{R}\!=\! 
\frac{1}{2}(\mathbb{I}\!+\!\boldsymbol{\pi})\psi$ as the two chiral projections we will eventually employ in the paper.

The inventory of all possible terms is quick: the least-order derivative model is the renormalizable model and it has contributions up to the $4$-dimensional mass terms
\begin{eqnarray}
\nonumber
&\mathcal{L}\!=\!A\frac{i}{2}
\left[\overline{\psi}_{L}\boldsymbol{\gamma}^{\mu}\boldsymbol{\nabla}_{\mu}\psi_{L}
-\boldsymbol{\nabla}_{\mu}\overline{\psi}_{L}\boldsymbol{\gamma}^{\mu}\psi_{L}\right]+\\
\nonumber
&+Z\frac{i}{2}\left[\overline{\psi}_{R}\boldsymbol{\gamma}^{\mu}\boldsymbol{\nabla}_{\mu}\psi_{R}
-\boldsymbol{\nabla}_{\mu}\overline{\psi}_{R}\boldsymbol{\gamma}^{\mu}\psi_{R}\right]+\\
\nonumber
&+C\overline{\psi}_{L}\boldsymbol{\gamma}^{\mu}\psi_{L} V_{\mu}
\!+\!H\overline{\psi}_{L}\boldsymbol{\gamma}^{\mu}\psi_{L} W_{\mu}+\\
\nonumber
&+S\overline{\psi}\boldsymbol{\gamma}^{\mu}_{R}\psi_{R} V_{\mu}
\!+\!X\overline{\psi}_{R}\boldsymbol{\gamma}^{\mu}\psi_{R} W_{\mu}-\\
&-\beta\overline{\psi}_{L}\psi_{R}-\beta^{\ast}\overline{\psi}_{R}\psi_{L}
\end{eqnarray}
where $A$, $Z$ and $C$, $H$, $S$, $X$ are real while $\beta$ is complex and all these parameters have still to be determined.

Because left-handed and right-handed semi-spinorial chiral projections must both have positive-defined energy then their kinetic terms must have the same sign, so that it is possible through a rescaling of $\psi_{L}$ and $\psi_{R}$ to set the parameters $A$ and $Z$ equal to unity and therefore we have
\begin{eqnarray}
\nonumber
&\mathcal{L}\!=\!
\frac{i}{2}\left(\overline{\psi}\boldsymbol{\gamma}^{\mu}\boldsymbol{\nabla}_{\mu}\psi
-\boldsymbol{\nabla}_{\mu}\overline{\psi}\boldsymbol{\gamma}^{\mu}\psi\right)+\\
\nonumber
&+K\overline{\psi}\boldsymbol{\gamma}^{\mu}\boldsymbol{\pi}\psi W_{\mu}
\!+\!F\overline{\psi}\boldsymbol{\gamma}^{\mu}\psi W_{\mu}+\\
\nonumber
&+E\overline{\psi}\boldsymbol{\gamma}^{\mu}\psi V_{\mu}
\!+\!J\overline{\psi}\boldsymbol{\gamma}^{\mu}\boldsymbol{\pi}\psi V_{\mu}-\\
&-m\overline{\psi}\psi\!-\!ib\overline{\psi}\boldsymbol{\pi}\psi
\end{eqnarray}
where $E$, $F$, $J$, $K$ and $m$ and $b$ are real parameters.

The non-completely antisymmetric irreducible tensorial part of torsion is absent and there is parity-invariance in the kinetic term although there is no definite parity in the potential terms, and this model has a least-order derivative and renormalizable Lagrangian.
\section*{CONSEQUENCES}
In the previous sections, we have discussed what happens when a model taking into account the torsional completion of gravitation with electrodynamics and Dirac fields is investigated under the requirement of continuity, finding two cases: one in which for torsion-gravity there are $2$-dimensional mass terms and for electrodynamics there are $4$-dimensional mass terms with completely antisymmetric torsion and parity-invariance, and for the Dirac matter field there are up to $4$-dimensional mass terms in the action; another in which for torsion-gravity as well as for electrodynamics there are $4$-dimensional mass terms with completely antisymmetric torsion and parity-invariance, and for the Dirac matter field there are up to $4$-dimensional mass terms in the action.

Then, because when the torsion is restricted to have complete antisymmetry in one sector of the theory so it is restricted in the entire theory, we will take torsion to be the completely antisymmetric dual of an axial vector, and as a consequence of this constraint all parity-odd terms in principle allowed in the $2$-dimensional mass model of torsion-gravity disappear: so the two models may be condensed together into the single Lagrangian as given by
\begin{eqnarray}
\nonumber
&\mathcal{L}\!=\!X(\nabla_{\alpha}W_{\nu}\!-\!\nabla_{\nu}W_{\alpha})
(\nabla^{\alpha}W^{\nu}\!-\!\nabla^{\nu}W^{\alpha})+\\
\nonumber
&+Y\nabla_{\alpha}W_{\nu}\nabla^{\alpha}W^{\nu}
\!+\!H\left|W_{\nu}W^{\nu}\right|^{2}+\\
\nonumber
&+SR_{\alpha\nu}W^{\alpha}W^{\nu}\!+\!URW_{\nu}W^{\nu}+\\
\nonumber
&+NR_{\alpha\mu}R^{\alpha\mu}\!+\!PR^{2}-\\
\nonumber
&-kR\!+\!BW_{\nu}W^{\nu}-\\
\nonumber
&-\frac{1}{4}F^{\alpha\nu}F_{\alpha\nu}+\\
\nonumber
&+\frac{i}{2}\left(\overline{\psi}\boldsymbol{\gamma}^{\mu}\boldsymbol{\nabla}_{\mu}\psi
-\boldsymbol{\nabla}_{\mu}\overline{\psi}\boldsymbol{\gamma}^{\mu}\psi\right)+\\
&+K\overline{\psi}\boldsymbol{\gamma}^{\mu}\boldsymbol{\pi}\psi W_{\mu}
\!+\!F\overline{\psi}\boldsymbol{\gamma}^{\mu}\psi W_{\mu}
\!-\!m\overline{\psi}\psi\!-\!ib\overline{\psi}\boldsymbol{\pi}\psi
\end{eqnarray}
where $N$, $P$, $S$, $U$, $X$, $Y$, $H$, $B$, $K$, $F$, $k$, $m$, $b$ are real, and for $N\!=\!P\!=\!S\!=\!U\!=\!X\!=\!Y\!=\!H\!=\!0$ we have the simplest least-order derivative Lagrangian in general.

This Lagrangian yields field equations whose consistency in terms of the amount of degrees of freedom and the character of propagation is to be checked with the method presented in \cite{Velo:1970ur} by Velo and Zwanziger.
\subsection{Consistent Propagation}
Just above we have given what, under the requirement of continuity, is the Lagrangian in its most general form
\begin{eqnarray}
\nonumber
&\mathcal{L}\!=\!X(\nabla_{\alpha}W_{\nu}\!-\!\nabla_{\nu}W_{\alpha})
(\nabla^{\alpha}W^{\nu}\!-\!\nabla^{\nu}W^{\alpha})+\\
\nonumber
&+Y\nabla_{\alpha}W_{\nu}\nabla^{\alpha}W^{\nu}
\!+\!H\left|W_{\nu}W^{\nu}\right|^{2}+\\
\nonumber
&+SR_{\alpha\nu}W^{\alpha}W^{\nu}\!+\!URW_{\nu}W^{\nu}+\\
\nonumber
&+NR_{\alpha\mu}R^{\alpha\mu}\!+\!PR^{2}-\\
\nonumber
&-kR\!+\!BW_{\nu}W^{\nu}-\\
\nonumber
&-\frac{1}{4}F^{\alpha\nu}F_{\alpha\nu}+\\
\nonumber
&+\frac{i}{2}\left(\overline{\psi}\boldsymbol{\gamma}^{\mu}\boldsymbol{\nabla}_{\mu}\psi
-\boldsymbol{\nabla}_{\mu}\overline{\psi}\boldsymbol{\gamma}^{\mu}\psi\right)+\\
&+K\overline{\psi}\boldsymbol{\gamma}^{\mu}\boldsymbol{\pi}\psi W_{\mu}
\!+\!F\overline{\psi}\boldsymbol{\gamma}^{\mu}\psi W_{\mu}
\!-\!m\overline{\psi}\psi\!-\!ib\overline{\psi}\boldsymbol{\pi}\psi
\end{eqnarray}
with $N\!=\!P\!=\!S\!=\!U\!=\!X\!=\!Y\!=\!H\!=\!0$ to obtain the simplest least-order derivative Lagrangian in general.

As a first thing, we have to notice that the number of independent fields and field equations must match.

Varying with respect to the axial vector torsion gives
\begin{eqnarray}
\nonumber
&2(2X\!+\!Y)\nabla^{2}W^{\nu}\!-\!4X\nabla^{\nu}\nabla_{\alpha}W^{\alpha}-\\
\nonumber
&-2(S\!+\!2X)R^{\alpha\nu}W_{\alpha}\!-\!2URW^{\nu}\!-\!4HW^{2}W^{\nu}-\\
&-2BW^{\nu}\!=\!F\overline{\psi}\boldsymbol{\gamma}^{\nu}\psi
\!+\!K\overline{\psi}\boldsymbol{\gamma}^{\nu}\boldsymbol{\pi}\psi
\end{eqnarray}
which is in fact a field equation because one may solve for the second-order time derivative of every component of the axial vector torsion, but its divergence is given by
\begin{eqnarray}
\nonumber
&2Y\nabla^{2}\nabla_{\alpha}W^{\alpha}\!-\!2[(S\!-\!Y)R^{\alpha\nu}
\!+\!4HW^{\alpha}W^{\nu}]\nabla_{\nu}W_{\alpha}-\\
\nonumber
&-2(UR\!+\!2HW^{2}\!+\!B)\nabla_{\nu}W^{\nu}-\\
&-(S\!-\!Y\!+\!2U)\!\nabla_{\alpha}RW^{\alpha}\!\!=\!\!\nabla_{\nu}
\!\left(F\overline{\psi}\boldsymbol{\gamma}^{\nu}\psi
\!+\!K\overline{\psi}\boldsymbol{\gamma}^{\nu}\boldsymbol{\pi}\psi\right)
\end{eqnarray}
which develops a third-order time derivative of the temporal component of the axial vector torsion, and therefore the system of field equations is not well defined, unless we require that the third-order derivative disappears, that is unless we ask $Y\!=\!0$ to hold; when this is done, we have that its divergence is reduced to the following expression
\begin{eqnarray}
\nonumber
&-2(SR^{\alpha\nu}\!+\!4HW^{\alpha}W^{\nu})\nabla_{\nu}W_{\alpha}-\\
\nonumber
&-2(UR\!+\!2HW^{2}\!+\!B)\nabla_{\nu}W^{\nu}-\\
&-(S\!+\!2U)\nabla_{\alpha}RW^{\alpha}\!=\!\!\nabla_{\nu}
\!\left(F\overline{\psi}\boldsymbol{\gamma}^{\nu}\psi
\!+\!K\overline{\psi}\boldsymbol{\gamma}^{\nu}\boldsymbol{\pi}\psi\right)
\label{c}
\end{eqnarray}
which has no third-order nor second-order time derivative of the temporal component of the axial vector torsion and therefore it is a true constraint, and although now the field equation is reduced to the following form
\begin{eqnarray}
\nonumber
&4X(\nabla^{2}W^{\nu}\!-\!\nabla^{\nu}\nabla_{\alpha}W^{\alpha})-\\
\nonumber
&-2(S\!+\!2X)R^{\alpha\nu}W_{\alpha}\!-\!2URW^{\nu}\!-\!4HW^{2}W^{\nu}-\\
&-2BW^{\nu}\!=\!F\overline{\psi}\boldsymbol{\gamma}^{\nu}\psi
\!+\!K\overline{\psi}\boldsymbol{\gamma}^{\nu}\boldsymbol{\pi}\psi
\label{fe}
\end{eqnarray}
which is not a true field equation since the second-order time derivative of the temporal component of the axial vector torsion never appears, nevertheless substituting in the field equation (\ref{fe}) the constraint (\ref{c}) we get that
\begin{eqnarray}
\nonumber
&4X\nabla^{2}W^{\nu}\!+\!4X[2(UR\!+\!2HW^{2}\!+\!B)]^{-1}\cdot\\
\nonumber
&\cdot\nabla^{\nu}[2(SR^{\alpha\mu}\!+\!4HW^{\alpha}W^{\mu})\nabla_{\mu}W_{\alpha}+\\
\nonumber
&+(S\!+\!2U)\nabla_{\alpha}RW^{\alpha}\\
\nonumber
&+\nabla_{\mu}\!\left(F\overline{\psi}\boldsymbol{\gamma}^{\mu}\psi
\!+\!K\overline{\psi}\boldsymbol{\gamma}^{\mu}\boldsymbol{\pi}\psi\right)]-\\
\nonumber
&-8X\nabla^{\nu}(UR\!+\!2HW^{2})\cdot\\
\nonumber
&\cdot[2(UR\!+\!2HW^{2}\!+\!B)]^{-2}\cdot\\
\nonumber
&\cdot[2(SR^{\alpha\mu}\!+\!4HW^{\alpha}W^{\mu})\nabla_{\mu}W_{\alpha}+\\
\nonumber
&+(S\!+\!2U)\nabla_{\alpha}RW^{\alpha}\\
\nonumber
&+\nabla_{\mu}\!\left(F\overline{\psi}\boldsymbol{\gamma}^{\mu}\psi
\!+\!K\overline{\psi}\boldsymbol{\gamma}^{\mu}\boldsymbol{\pi}\psi\right)]-\\
\nonumber
&-2(S\!+\!2X)R^{\alpha\nu}W_{\alpha}\!-\!2URW^{\nu}\!-\!4HW^{2}W^{\nu}-\\
&-2BW^{\nu}\!=\!F\overline{\psi}\boldsymbol{\gamma}^{\nu}\psi
\!+\!K\overline{\psi}\boldsymbol{\gamma}^{\nu}\boldsymbol{\pi}\psi
\label{tfe}
\end{eqnarray}
which is a true field equation because the second-order time derivative of every component of the axial vector torsion is present: therefore $Y\!=\!0$ is a constraint which ensures that the amount of the physical degrees of freedom and the number of the independent field equations correspond precisely as consistency arguments dictate.

To see what happens about the propagation, we have to consider only the highest-order derivatives, and after the substitution $i\nabla_{\alpha}\rightarrow n_{\alpha}$ the characteristic equation is
\begin{eqnarray}
\nonumber
&(UR\!+\!2HW^{2}\!+\!B)n^{2}\!+\!SR^{\alpha\nu}n_{\alpha}n_{\nu}+\\
&+4H|W^{\alpha}n_{\alpha}|^{2}\!=\!0
\label{ce}
\end{eqnarray}
and field equations (\ref{tfe}) cease to be causal whenever the characteristic equation (\ref{ce}) allows $n^{2}\!>\!0$ to occur.

In order to study in what way the characteristic equation may affect the causal propagation, we will consider its limiting cases, the first of which being the case in which torsion is small, and in which also curvature is small, so that the characteristic equation becomes
\begin{eqnarray}
&Bn^{2}\!+\!SR^{\alpha\nu}n_{\alpha}n_{\nu}\!\approx\!0
\end{eqnarray}
and as we have no information about $R^{\alpha\nu}$ acausality may occur, unless $S\!=\!0$ holds as constraint; but even in this case, in the same approximation in which torsion is small, but in the complementary approximation in which curvature is large, the characteristic equation becomes
\begin{eqnarray}
&URn^{2}\!+\!4H|W^{\alpha}n_{\alpha}|^{2}\!\approx\!0
\end{eqnarray}
and as we have no information about $R$ acausality may occur, unless $U\!=\!0$ holds as constraint; but again even in this case, in the complementary approximation in which torsion is large, the characteristic equation becomes
\begin{eqnarray}
&2HW^{2}n^{2}\!+\!4H|W^{\alpha}n_{\alpha}|^{2}\!\approx\!0
\end{eqnarray}
and because the axial vector torsion cannot have only one degree of freedom then $W^{2}$ cannot be time-like and acausality occurs, unless $H\!=\!0$ holds; hence, the characteristic equation reduces to $n^{2}\!=\!0$ and then the causality is preserved: so $S\!=\!U\!=\!H\!=\!0$ is a set of constraints which ensures that the causal propagation is maintained.

Furthermore, the torsion-spin coupling equations have a parity symmetry for which $F\!=\!0$ and for the whole Lagrangian the same symmetry gives $b\!=\!0$ identically.

We may finally summarize: for renormalizable models torsion is dynamical and relevant at all scales, so that there are field equations whose consistent propagation requires $Y\!=\!S\!=\!U\!=\!H\!=\!F\!=\!b\!=\!0$ as the most stringent constraints that are possible, and so that the remaining constraints given by $N\!=\!P\!=\!X\!=\!0$ define the least-order derivative Lagrangian in the most general case.
\section*{EFFECTS}
In this discussion, we have seen that the most general Lagrangian is reduced to the one given by the following
\begin{eqnarray}
\nonumber
&\mathcal{L}\!=\!X(\nabla_{\alpha}W_{\nu}\!-\!\nabla_{\nu}W_{\alpha})
(\nabla^{\alpha}W^{\nu}\!-\!\nabla^{\nu}W^{\alpha})+\\
\nonumber
&+NR_{\alpha\mu}R^{\alpha\mu}\!+\!PR^{2}-\\
\nonumber
&-kR\!+\!BW_{\nu}W^{\nu}-\\
\nonumber
&-\frac{1}{4}F^{\alpha\nu}F_{\alpha\nu}+\\
\nonumber
&+i\overline{\psi}\boldsymbol{\gamma}^{\mu}\boldsymbol{\nabla}_{\mu}\psi+\\
&+K\overline{\psi}\boldsymbol{\gamma}^{\mu}\boldsymbol{\pi}\psi W_{\mu}
\!-\!m\overline{\psi}\psi
\label{l}
\end{eqnarray}
where $N\!=\!P\!=\!X\!=\!0$ are the constraints that define the most general least-order derivative Lagrangian possible.

In general this Lagrangian is renormalizable \cite{s}.

Therefore, as it follows from this discussion, the property of renormalizability does not need to be arbitrarily assumed because it can be directly derived by insisting that parameters $N$, $P$ and $X$ be different from zero in the most general circumstances; the least-order derivative Lagrangian for $N\!=\!P\!=\!X\!=\!0$ can be seen as low-energy approximation of the general one: the difference in the two cases is that at high-energy regimes, the least-order derivative model will always remain an effective model with no associated torsion boson, while in the renormalizable model a massive neutral axial-vector torsion boson will have to be expected. No such torsion boson has ever been detected, which constitutes an indication against the assumption of renormalizability; this situation is not so bad since in presence of torsion for dynamics in non-minimal coupling the effective renormalizability can be recovered, as discussed in \cite{Fabbri:2014wda}: this circumstance seems to point out that whenever torsion is 
present in gravity the concept of field renormalization might have to be rethought altogether. In fact, the requirement that gravity must be relevant even at extremely small scales might well be a prejudice coming from the fact that we think at gravity as a force like any other force, while clearly this is not the case; being gravity a manifestation of a property of the underlying geometric background for which it tends to disappear locally, it is all too reasonable that the common arguments about ultra-violet renormalizability cannot be applied. Because renormalizability means that the kinetic term of a field must never be suppressed by some interaction of that field then if this were possible it would mean that the field would tend to vanish, which is exactly what happens for gravity whenever the gravitational field is considered to be the spacetime curvature.

As a property for which some term ought be relevant at all scales presupposes the knowledge of a physics that is either against evidence as in the case of gravity or precluded altogether, renormalizability cannot be considered as a fundamental principle: this is not a problem, because effectively it can be obtained in terms of generality arguments, as we have shown here; alternatively, if there is no renormalizability then we may have only the least-order derivative model, which can be obtained by simply imposing the requirement of dealing with the least order of the differential structure as a fundamental principle.
\section*{CONCLUSION}
In the present paper, we have studied torsion-gravity with electrodynamics and Dirac fields and we have seen that arguments of continuity in the torsionless limit and consistency in time evolution and causal propagation justify the fact that torsion had to be the completely antisymmetric dual of an axial vector and that parity-conservation had to be a feature of the dynamics in a model described by the Lagrangian (\ref{l}) and where the special constraint $N\!=\!P\!=\!X\!=\!0$ defines the least-order derivative Lagrangian; in the end, we have given a general discussion about the meaning of renormalizability in the context of a theory that is geometrically constructed.

We have not discussed any of the consequences that could derive for a quantum theory of gravity.


\begin{thebibliography}{40}
\bibitem{Capozziello:2001mq}
S.Capozziello, G.Lambiase, C.Stornaiolo,\\
\textit{Annalen Phys.}\textbf{10}, 713 (2001).
\bibitem{a-l}
J.Audretsch, C.L\"{a}mmerzahl,\\
\textit{Class. Quant. Grav.} \textbf{5}, 1285 (1988).
\bibitem{m-l}
C.L\"{a}mmerzahl, A.Macias,\\
\textit{J.Math.Phys.}\textbf{34}, 4540 (1993).
\bibitem{Fabbri:2006xq}
L.Fabbri,
\textit{Ann.Fond.Broglie}\textbf{32}, 215 (2007).
\bibitem{Fabbri:2009se} 
L.Fabbri,
in \textit{Einstein and Hilbert: Dark\\ Matter} (Dvoeglazov, Nova Publishers, 2012).
\bibitem{Fabbri:2008rq} 
L.Fabbri,
\textit{Annales Fond. Broglie}\textbf{33}, 365 (2008).
\bibitem{Fabbri:2009yc} 
L.Fabbri,
\textit{Int. J. Theor. Phys.}\textbf{51}, 954 (2012).
\bibitem{FV}
L.Fabbri, S.Vignolo,
\textit{Int.J.Theor.Phys.} \textbf{51}, 3186 (2012).
\bibitem{Fabbri:2011kq} 
L.Fabbri,
\textit{Gen.Rel.Grav.}\textbf{45}, 1285 (2013).
\bibitem{Fabbri:2013gza} 
L.Fabbri,
\textit{Gen.Rel.Grav.}\textbf{46}, 1663 (2014).
\bibitem{s-s}
C.Sivaram, K.P.Sinha,
\textit{Lett. Nuovo Cim.} \textbf{13}, 357 (1975).
\bibitem{Fabbri:2010rw}
L.Fabbri,
\textit{Mod.Phys.Lett.A}\textbf{27}, 1250028 (2012).
\bibitem{Fabbri:2012ag} 
L.Fabbri,
\textit{Int.J.Theor.Phys.}\textbf{53}, 1896 (2014).
\bibitem{Fabbri:2013isa} 
L.Fabbri,
\textit{Int.J.Theor.Phys.} \textbf{53}, 3744 (2014).
\bibitem{Fabbri:2014naa} 
L.Fabbri,
\textit{Int.J.Geom.Meth.Mod.Phys.}\textbf{11},1450073(2014).
\bibitem{Fabbri:2014zya} 
L.Fabbri,
\textit{Gen.Rel.Grav.}\textbf{47}, 1837 (2015).
\bibitem{Bender2008/1}
C.M.Bender, P.D.Mannheim,\\
\textit{Phys.Rev.Lett.}\textbf{100}, 110402 (2008).
\bibitem{Bender2008/2}
C.M.Bender, P.D.Mannheim,\\
\textit{Phys.Rev.D}\textbf{78}, 025022 (2008).
\bibitem{Fabbri:2014kea}
L.Fabbri, P.D.Mannheim,
\textit{Phys.Rev.D}\textbf{90}, 024042 (2014).
\bibitem{Velo:1970ur} 
G.Velo, D.Zwanziger,
\textit{Phys. Rev.}\textbf{188}, 2218 (1969).
\bibitem{s}
K.~S.~Stelle,
\textit{Phys. Rev. D} \textbf{16}, 953 (1977).
\bibitem{Fabbri:2014wda} 
L.Fabbri, S.Vignolo, S.Carloni,\\
\textit{Phys.Rev.D}\textbf{90}, 024012 (2014).
\end{thebibliography}
\end{document}